A STUDY OF AN ACRYLIC CERENKOV RADIATION DETECTOR

B. Porter, P. Auchincloss, P. de Barbaro, A. Bodek, and H. Budd

Department of Physics and Astronomy, University of Rochester, Rochester, New York 14627

**INTRODUCTION**

An experiment investigating the angle of Cerenkov light emitted by 3-MeV electrons traversing an acrylic detector has been developed for use in the advanced physics laboratory course at the University of Rochester. In addition to exploring the experimental phenomena of Cerenkov radiation and total internal reflection, the experiment introduces students to several experimental techniques used in actual high energy and nuclear physics experiments, as well as to analysis techniques involving Poisson statistics.

The interactions of very high energy charged particles with matter are usually investigated using high-intensity beams generated by particle accelerators. Most undergraduate students, however, do not have access to beams from particle accelerators. Sources of relativistic particles which are readily available in undergraduate teaching labs include primarily cosmic ray muons and electrons from radioactive beta decay. While the rate of cosmic ray muons is very low and their energy spectrum is wide, limiting their usefulness as a source of particles, the rate of high energy electrons from radioactive sources can be high. The experimental apparatus described here generates an intense beam of monoenergetic electrons using the combination of a radioactive source and a small permanent magnet beam line. This "table-top" experiment enables students to study fundamental interactions of high energy particles with matter, including such processes as Cerenkov radiation, *dE/dx* energy loss in matter, and multiple Coulomb scattering. Moreover, the apparatus teaches several basic techniques of experimental particle physics paralleling the techniques used in "real" experiments at major particle accelerator facilities. In this communication, we describe the construction and use of a table-top electron beam line in the study of Cerenkov radiation. An appendix contains supplier and cost information for apparatus components.

**A REVIEW OF CERENKOV RADIATION AND CALCULATION OF EXPECTED PHOTON YIELD**

The electric field of a charged particle moving in a given medium can propagate no faster than that medium's speed of light ($c/n$). If a charged particle moves with velocity v greater than the medium's speed of light, the particle appears to emit a "shock-wave" of light at an angle to the particle's trajectory, as depicted in Fig. 1(a). Fig. 1(b) indicates, for an electron, how to determine the characteristic Cerenkov angle θ. After a time *t*, the particle has moved a distance v*t*, and the electromagnetic field has traveled a distance of *ct*/*n*. Wavelets (as conceptualized in Huygen's principle) form at each point along the trajectory and additively produce a wavefront at right angles to the emitted photons. The angle (or, more precisely, the half-angle) of Cerenkov light θ is determined by the relation

$$\cos\theta = \frac{c}{vn} = \frac{1}{\beta n}, \tag{1}$$

where $\beta = v/c$. This formula implies that Cerenkov light is produced only when v > *c/n* and allows the expected Cerenkov cone angle to be predicted if v (or $\beta$) and *n* are known. Using the mass-energy relationship $E^2 = p^2c^2 + m^2c^4$ we can find the electron's $\beta$. In the case of a 3-MeV electron, for instance,

$$\beta = [1-(m_e/E)^2]^{1/2} = [1-(0.511/3)^2]^{1/2} = 0.985. \tag{2}$$

Thus, for *n* = 1.49 and $\beta$ = 0.985, the Cerenkov angle is θ = 47.0°.

The approximate number of visible-spectrum photons produced by the charged particle, as it traverses the length (*L*) of the dielectric medium, is given by[1]

$$N_p = 400L(\sin\theta)^2. \tag{3}$$

For a material of thickness 0.156 cm and θ = 47.0°, the approximate number of visible photons is 34. Such low levels of Cerenkov light requires the use of a photo-multiplier tube (PMT) to detect the photons emitted by the electron. In this experiment, a polished piece of acrylic is used to direct the photons into the PMT by total internal reflection. The PMT used is sensitive only to visible wavelengths, so Eq. (3) is valid.

Starting with 34 photons and an average quantum efficiency of 5% (over the entire visible range of a typical bi-alkali PMT), the expected number of photo-electrons exiting the photocathode is about 1.7. The amount of detectable light will be smaller if additional losses are incurred: losses due to light attenuation in the acrylic (about 30%, depending on the path-length to the PMT, or about 13.7 cm for this setup); losses at the interface between the PMT and the detector (about 20%); and losses due to poor optical machining of the surfaces. Therefore, if the entire angular cone of Cerenkov light is within the geometrical acceptance of the PMT, the expected number of photo-electrons is about 1.0. If the PMT does not accept the entire Cerenkov angular cone, the number of photo-electrons will be smaller.

**EXPERIMENTAL SETUP**

There are five components to this experiment: the 3 MeV Ruthenium beta source beam line[2], the acrylic Cerenkov radiation detector and PMT, the triggering system, the CAMAC computer interface[3] and software, and the light-tight box. Each of these components is discussed in the following subsections. A functional schematic of the setup is shown in Fig. 2.

**Ruthenium beta source and beam line**

The ruthenium (atomic number 44) isotope of atomic weight 106 (Ru-106) emits electrons via beta decay with an energy spectrum up to 3.541 MeV. In the experimental setup, electrons emitted by the Ru-106 source enter a beam line[2] comprised of two dipole and three quadrupole permanent magnet sections. By forcing the emitted electrons to bend through an angle (dependent on electron momentum) and selecting those which bend at 90-degrees, the beam line effectively selects only 3-MeV (±5%) electrons and directs them toward the scintillation counters. For a 4-millicurie Ru-106 beta source, the beam line delivers about 50 electrons per second. The beam line can be used with or without a vacuum to reduce multiple scattering. No vacuum was used in this experiment.

**Acrylic Cerenkov radiation detector**

The detector is a polished bar of acrylic or poly-methyl-methacralate (PMMA), with dimensions 32.5 cm x 5.0 cm x 0.156 cm. In developing this experiment, we have used an ultraviolet-transmitting acrylic (UVT), but because the PMT is not UV-sensitive, an ultraviolet-absorbing acrylic (UVA) could be used instead. As shown in Fig. 3(a), one end of the acrylic bar is glued into a mounting disk (called a "cookie"), while the other end is painted black and wrapped in black tape to reduce the amount of spurious light reflected into the PMT. The acrylic bar is covered with black plastic to prevent scratches and then secured to a metal support to prevent it from bending under its weight [Fig. 3(b)]. The support and cookie are attached with tape to a PMT [Fig. 4(a)], and a magnetic shield is placed around the PMT [Fig. 4(b)].

The output signal of the PMT is generated by applying a high voltage to the anode and cathode of the PMT. The fourteen stage PMT (Amperex 56 AVP) is biased at 1.8 kV and has a maximum voltage limit of 2.5 kV.

Acrylic has an index of refraction of 1.49 for visible light[4]. Because the refractive index of air is 1, total internal reflection occurs when the angle of the incident light ray is equal to or greater than 42.2°. When a relativistic electron ($v \approx c$) passes through the acrylic bar, it produces a cone of light, due to the Cerenkov effect. The light intensity profile of the cone of light can be mapped by varying the angle of the acrylic bar in relation to the path of the electron beam. For different angles, a larger or smaller fraction of the Cerenkov light may be captured by total internal reflection and, hence, detected by the PMT. This effect is illustrated in Fig. 5.

**Event trigger**

An "event" occurs when an electron passes through the Cerenkov detector and the scintillation counters. Each scintillation counter is comprised of a plastic scintillating plate and a PMT, as shown in Fig. 2. The PMT output signal is carried by a coaxial cable into a discriminator. The (nearly) simultaneous detection of an electron in both scintillation counters generates an event trigger signal which activates the computerized data acquisition system.

Figure 6 shows the electronic components and logic used to generate an event trigger signal. Two discriminators convert the top and bottom PMT signals into NIM logic signals (-0.7 or 0 V). When the PMT output voltage falls below a preset value, the discriminator sends out a -0.7 V pulse for a preset duration (typically 20 nanoseconds). The top and bottom discriminators connect to a logical "AND" gate that generates a signal when both top and bottom signals are nonzero. A successful "AND" signal generates the trigger, indicating the coincident firing of the top and bottom discriminators and detection of an electron in both scintillation counters. The trigger signal "turns on" the computerized data acquisition system by activating an analog-to-digital converter (ADC) that digitizes the signal from the Cerenkov PMT for a preset time duration or "trigger pulse duration." The trigger pulse duration is chosen to maximize the signal-to-noise ratio of the converted signal; a duration of 70 nsec was used in developing this experiment.

Because the transit time of the signal in each PMT may be different by a few nanoseconds, it is necessary to delay the signals appropriately so that all signals reach the logic electronics at the same time. To account for internal delay, the cable for the top counter is longer than the cable for the bottom counter. This allows the signals from both counters to reach the "AND" gate at the same time. Likewise, the Cerenkov PMT signal is delayed to allow the signal to arrive within the ADC gatewidth (that is, the trigger pulse duration). After the cable lengths have been properly selected, the experiment is "timed-in" and ready to take data.

Once analog-to-digital conversion of the Cerenkov PMT signal is complete, the digitized level of electric charge (corresponding to the amount of light captured in the PMT) is recorded in a computerized data file for analysis. This experiment uses a LeCroy 2249A ADC with a resolution of 10 bits, or 1024 distinct charge levels. The amount of Cerenkov light for a detected event is represented in the data file by a number from 0 to 1023, corresponding to the ADC charge level.

**Software**

A simple data acquisition program has been written in Turbo Pascal for an IBM compatible computer with CAMAC 6001 interface card. The program, called CERENKOV, uses standard CAMAC interface commands to read data from the ADC module on the CAMAC crate. A user can either collect data for a specified number of events or for a specified time period. The data files may be viewed with any text editor and may be analyzed to determine the number of detected photo-electrons. The documentation for the CAMAC card describes the command syntax for different programming languages.

**Light-tight box**

The light-tight box contains the controls, equipment, and radioactive source necessary to test the acrylic Cerenkov detector and scintillation counters. Here, "light-tight" means that no light can enter the box when the lid is latched shut. This feature protects the PMT's from high current damage while the high voltage power supply is energized.

As mentioned earlier, the Ru-106 radioactive source emits beta radiation (electrons) with an energy spectrum up to 3.5-MeV, and the subsequent beam line structure[2] selects 3-MeV electrons and directs them down through the scintillation counters. A support placed within the box holds the Cerenkov detector at a fixed angle to the electron beam. As shown in Fig. 7, a ring bolt allows the experimenter to vary the angle of the detector in relation to the beam. The detector-to-beam angle $\alpha$ can be measured with a plumb line and protractor.

**EXPERIMENTAL PROCEDURE**

The objective of this experiment is for the student to determine how the amount of Cerenkov light collected in the PMT varies with the detector-to-beam angle, and to compare the result to predictions based on theory and information about the apparatus. To do this, the student collects data in sets or "runs," each of which corresponds to a different detector-to-beam angle $\alpha$.

For each data run, 100 to 1000 events (described in the previous section) are recorded in a computer file. By using methods described in the next section, the student can determine the average number of photo-electrons per event trigger, which is proportional to the amount of Cerenkov radiation. Finally, the student prepares a plot of the average number of photo-electrons versus the detector-to-beam angle $\alpha$.

**DATA ANALYSIS**

The following analysis is used to calculate the number of photo-electrons detected by the Cerenkov PMT. As stated in the "Review of Cerenkov Radiation" above, the expected number of photo-electrons is about 1.0 if the entire angular cone of Cerenkov light is within the geometrical acceptance of the PMT. If the PMT does not accept the entire Cerenkov angular cone, the number of photo-electrons will be smaller. The first step is to distinguish pedestal from non-pedestal events. The ADC "pedestal" refers to the "zero" level of the detected light and corresponds to the ADC level (i.e., 0-1023) with the largest number of events. For a Poisson distribution, if the average number of photo-electrons is N, then $e^{-N}$ is the probability that no photo-electrons are detected. The number of photo-electrons N and the error are defined in Eqs. (4) and (5), respectively:

$$N = \#\text{of p.e.} = -\ln\frac{\#\text{ of events in pedestal}}{\text{total }\#\text{ of events}} \qquad (4)$$

$$N = \text{error (\# of p.e.)} = \left[\frac{\text{\# of p.e.}}{\text{total \# of events}}\right]^{\frac{1}{2}} \qquad (5)$$

The number of photons produced is proportional to the length of the electron's path in the acrylic bar. As the detector-to-beam angle ($\alpha$) increases, the electron's path-length also increases as $1/\cos(\ )$. Consequently, the number of photo-electrons must be corrected to account for the longer path-length and corresponding addition of photons. This is accomplished by multiplying Eq. 4 by $\cos(\ )$.

$$\text{\# of p.e. (corrected)} = -\ln\left[\frac{\text{\# of events in pedestal}}{\text{total \# of events}}\right] \cos(\alpha). \qquad (6)$$

To improve the accuracy of the experiment and provide a more realistic learning experience regarding high energy physics experiments, students may be asked to correct the number of photo-electrons due to background and other effects, as described below.

**Cosmic ray background**

The experimenter may investigate background events from cosmic rays (muons coming from the upper atmosphere). To study the cosmic ray background, it is necessary to record two sets of data for each detector-to-beam angle. For the first data set, the experimenter places the detector beneath the electron beam line and collects data normally for a fixed amount of time. For the second data set, the experimenter blocks the electron beam line with a lead plate while collecting data. Because the lead plate blocks the electrons from the Ru-106 source, the scintillation counters trigger only on cosmic ray events. The ADC spectrum produced by cosmic rays can be subtracted from the normal spectrum (which results from the sum of Cerenkov and cosmic ray events) before estimating the number of photo-electrons.

**Multiple scattering**

In the above formulae, the electron's path through the acrylic is assumed to be straight. Multiple scattering, however, may cause the electron to deviate from the original trajectory. This deviation tends to blur the angular dependence of Cerenkov light, so it is important to calculate the expected angular deviation for an electron. The average root-mean-square (rms) multiple scattering angle[5] in radians is given by Eq. (7):

$$\phi_{rms} = \frac{z 21 \text{MeV}}{pv} \left(\frac{t}{X_0}\right)^{\frac{1}{2}}, \qquad (7)$$

where $z$ is charge of the particle, $p$ is the particle's momentum, v is the particle's velocity, $t$ is the thickness of the material through which the particle is moving, and $X_0$ is the radiation length of material. Eq. (7) gives the total rms angle in space for multiple scattering in the x and y planes. If a 3-MeV electron (v/c close to 1.0) travels through acrylic, the rms multiple scattering angle is 27°, given $t = 0.1563$ cm, $X_0 = 34.4$ cm, $z = 1$, and $pv = 3$ MeV (see reference 6). This rms angular deviation is about half of the Cerenkov angle (about 47°), which would cause considerable error in the measurement. Fortunately, the trigger system selects well-collimated electrons with actual deviations of only a few degrees or those electrons which have scattered out and then back into the beam line.

**Spurious photons reflected into the PMT**

Black paint and tape are used to prevent reflected light from reaching the PMT, as shown in Fig. 3. However, the present detector configuration may not stop all reflections. A better configuration is to cut the end of the acrylic bar at a 45° angle and then paint it black, as shown in Fig. 8.

**EXPERIMENTAL RESULTS**

This experiment has been conducted for positive and negative detector-to-beam angles ($\alpha$), and the Cerenkov light profile is plotted in Fig. 9. The amount of light detected by the PMT can be seen to vary with the angle of the acrylic bar. For a 3-MeV electron, the calculated Cerenkov angle θ is 47.0°. Due to the acrylic-air interface, the critical angle for total internal reflection is 42.2°. For normal incidence, the Cerenkov photons traveling in the direction of the PMT would be trapped in the bar and detected. If the incident angle is allowed to change, there will be an angle for which total internal reflection does not trap the Cerenkov light in the direction of the PMT. If any light directed away from the PMT is trapped by total internal reflection, it is then absorbed by the black paint at the end of the bar. This total internal reflection cutoff angle is approximately -5°. This means that the number of photo-electrons detected should increase as α is increased, and no light should be detected for α < -5°.

The general shape of the graph conforms to this expectation, but light is detected for α < -5°. Several factors may account for the detection of light even when the detector-to-beam angle α is less than the cutoff for total internal reflection. First, the experimental model does not account for blemishes and scratches on the acrylic bar. Scratches cause scattering of photons, potentially destroying the angular dependence of the Cerenkov radiation. Second, reflection of light from the edges of the acrylic bar could cause light to be detected for α < -5°. Third, if the bar of acrylic were to bend under its own weight, the angular dependence of the Cerenkov light again would be affected. (We note that the data reported here were taken with an early version of the experiment; the current setup includes support structures to prevent sagging.) Fourth, cosmic ray background could produce detectable light for angles < -5°. Fifth, multiple scattering could result in small deviations in angular dependence. It is left to the student to investigate these possible sources of background and try to improve the experimental setup.

Several further studies are possible with this experimental setup:

1. Students may estimate the expected number of photo-electrons more precisely and derive the average quantum efficiency of the PMT.

2. After determining the angle which maximizes the Cerenkov light yield, students may measure the attenuation length of the light in acrylic by varying the distance between the point at which the electron beam crosses the acrylic bar and the PMT.
3. Students may write a Monte Carlo simulation program, which includes multiple scattering and other effects, to reproduce the plot of detected light versus angle.
4. The experiment may be modified by replacing the acrylic bar with a thin scintillator. A thin scintillation counter yields a large number of photo-electrons for a minimum ionizing particle traversing the scintillator. Data analysis could then include a study of Landau energy-loss by electrons and cosmic ray muons passing through matter.

**CONCLUSION**

The Cerenkov effect is a fascinating physical phenomenon which provides an important experimental technique for the detection and counting of high energy, charged particles. The experiment described here engages students with relativistic kinematics, optical phenomena, scintillation light, photo-electron production, and the Cerenkov effect.

In addition, students are introduced to instruments and techniques frequently used in high energy experiments. The electron beam line replicates, on a small scale, the beam lines used at major particle accelerators. The relatively high rate of beta production from the Ru-106 source (50 electrons per second) allows students to "time-in" the fast-logic electronics on an oscilloscope, perform timing studies, and run a variety of systematic checks. The students work with electronic logic circuits, photo-multiplier tubes, and computerized data acquisition systems. The data analysis requires Poisson statistics and an investigation of systematic errors, which in turn deal with fundamental physical processes and ordinary physical phenomena such as multiple scattering and cosmic rays. The experiment may be extended to include studies of the quantum efficiency of photo-multiplier tubes and Landau energy loss processes. The experiment is also amenable to computerized simulation, allowing closer examination of many of the phenomena it presents.

For additional information on Cerenkov radiation, see *Cerenkov Radiation in High-Energy Physics: Part II, Cerenkov Counters*, edited by V. P. Zrelov and published by Keter Press, Jerusalem, Israel, 1970.

**APPENDIX**

The electron beam line was designed according to the general specifications described in reference 2. Figure 10 shows a side and bottom view of the beam line as it was designed at Rochester. The beam line uses 3 quadrupole magnet structures (comprised of 8 magnets) and 2 dipole magnet structures (comprised of 6 magnets). The size and shape of the ceramic magnets are shown in Figure 11. A necessary design feature is the ability to adjust the magnet positions around the beam line. This allows one to better focus the electron beam when assembling the beam line from commercial permanent magnets with small variations in field strength. The bending magnets are attached to the mounting plates with screws which go into radial slots. The quadrupole magnets are assembled around the beam pipe, and are therefore movable in the longitudinal direction along the beam line.

The electron beam line traverses a 90 degree bend which is approximately circular. The bending magnet active area is triangular, which can be used to change the path-length of the electrons in the magnetic field. If the magnet is moved to a larger radius, the electrons traverse a shorter length in the magnetic field, thus selecting a higher electron beam energy. Selection of a higher beam energy results in a lower rate of electrons

There are two methods to construct the beam line. In the first method, one can either measure the field of each magnet and position the magnet according to the desired beam energy. This method is time consuming and required the use of a Hall probe. If knowledge of the precise beam energy is not needed, one can fix the radial location of the first bending magnet, and vary the radial location of the second magnet until the beam rate is maximized. If the beam rate is measured using a radiation monitor or a Geiger counter, the beam line can be set up in about 10 minutes.

It is assumed that most laboratories have a computer which can be used for the data acquisition system. The CAMAC system that has been used in our setup is not specifically required, and any computer with a standard analog to digital conversion card will work. The software used in our setup is available on the WWW at *http://www.pas.rochester.edu/ug.html* under the "Senior Lab Experiment Notes" link.

The following section lists the cost and supplier for major (non-standard) components of this experiment.

**Radioactive source and beam line:**

| | | |
|---|---|---|
| 4-millicurie, Ru-106 source | $2,245 | North American Scientific<br>(818) 503-9201 |
| Dipole and quadrupole magnets | $880 | Adam's Magnetic Products, CO.<br>2081 N. 15$^{th}$ Ave.<br>Melrose Park, IL 60160 |

**Detector:**

| | | |
|---|---|---|
| Photo-multiplier tube w/base and shield | $410 | Hamamatsu Corporation<br>360 Foothill Rd.<br>Bridgewater, NJ 08807 |
| Acrylic thin plate | $10 | Supplier: any acrylic/plastic reseller. |

**ACKNOWLEDGMENTS**

This work has been supported in part by the Research Experiences for Undergraduates Site Program of the National Science Foundation, Grant No. PHY-9400059, and the University of Rochester.

**REFERENCES**

[1] A. Melissinos, *Experiments in Modern Physics* (Academic Press, New York, NY, 1966) p 174.

[2] David G. Underwood, U.S. Patent Number 5,198,674; David G. Underwood, Argonne National Laboratory, private communication; 1997 (to be published in *Nucl. Inst. Meths.*)

[3] W. R. Leo, *Techniques for Nuclear and Particle Physics Experiments* (Springer-Verlag, Berlin, Germany, 1987), pp. 327-341. See also LeCroy product catalog.

[4] M. J. Weber, ed., *CRC Handbook of Laser Science and Technology, Volume IV, Optical Materials Part 2* (CRC Press, Inc., Boca Raton, FL, 1986), pp. 85-91.

[5] D. H. Perkins, *Introduction to High Energy Physics* (Addison-Wesley Publishing Company, Melno Park, CA, 1987), 3rd ed., p 41.

[6] The radiation length for acrylic was obtained from the *Review of Particle Properties, Phys. Rev.* D50, 1173 (1994), p 1241.

**FIGURE CAPTIONS**

Fig. 1. (a) Diagram showing the production of Cerenkov light, where v is the particle velocity, *c* is the speed of light in vacuum, and *n* is the dielectric refractive index. *c/n* is the speed of light in the dielectric. (b) Graphical derivation of $\theta$.

Fig. 2. Experimental setup for the detection of Cerenkov radiation, side view.

Fig. 3. Schematic view of (a) acrylic Cerenkov detector and (b) detector with metal support.

Fig. 4. The acrylic Cerenkov detector is shown attached to (a) photo-multiplier tube and (b) photo-multiplier tube with magnetic shield.

Fig. 5. Cerenkov detector (side view) showing different incident angles: (a) $\alpha = 0°$, (b) $\alpha = 20°$, (c) $\alpha = -20°$.

Fig. 6. Block diagram of electronic components and logical inter-connections.

Fig. 7. Schematic view of adjustable support: (a) back view and (b) side view.

Fig. 8. Improvements to experimental setup, side view.

Fig. 9. Graph of photo-electrons (as calculated in text) vs. detector-to-beam angle $\alpha$.

Fig. 10. Schematic of beam line (a) side view and (b) bottom view

Fig. 11. Dimensions for (a) dipole and (b) quadrupole magnets, where the arrows indicate the direction of the magnetic field.

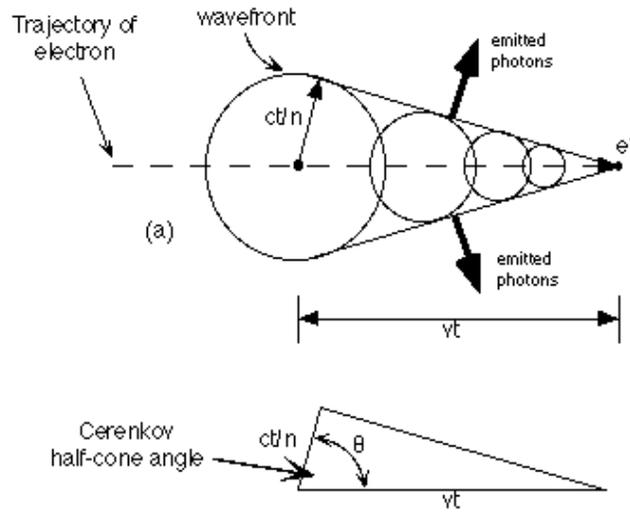

Fig. 1
Brian Porter
*American Journal of Physics*

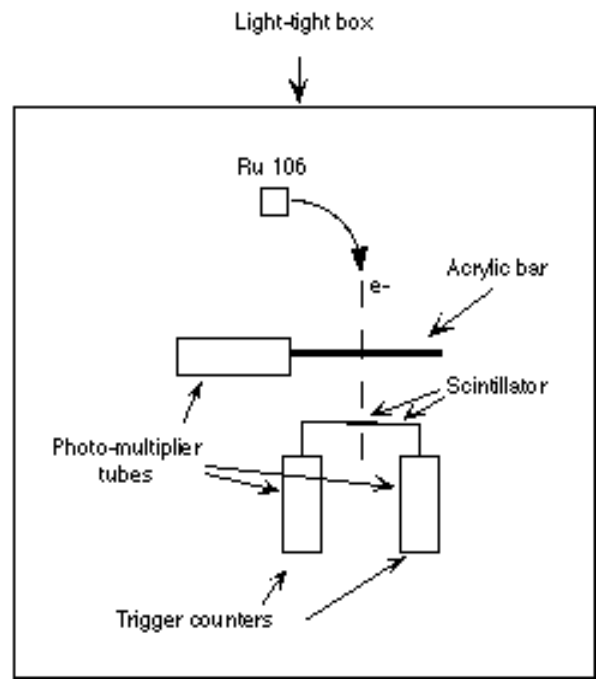

Fig. 2
Brian Porter
*American Journal of Physics*

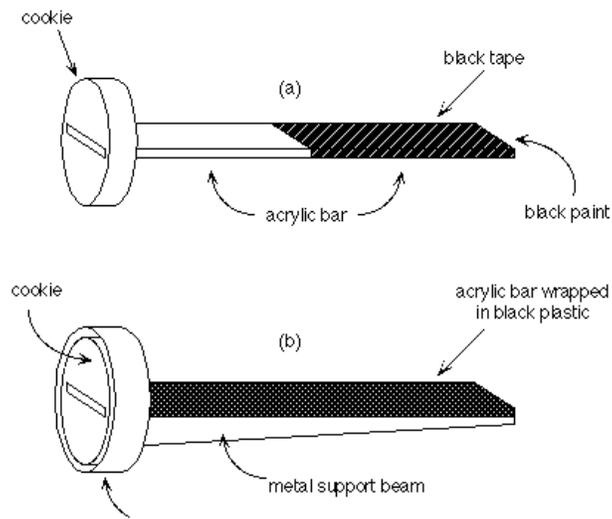

Fig. 3
Brian Porter
*American Journal of Physics*

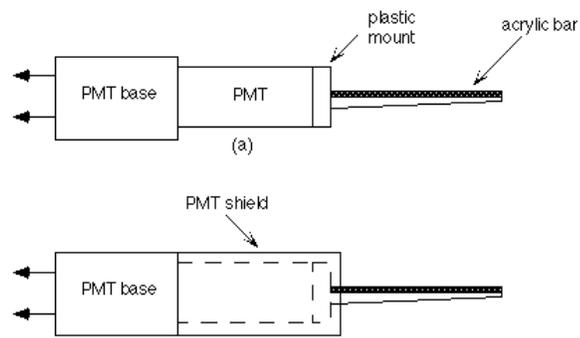

Fig. 4
Brian Porter
*American Journal of Physics*

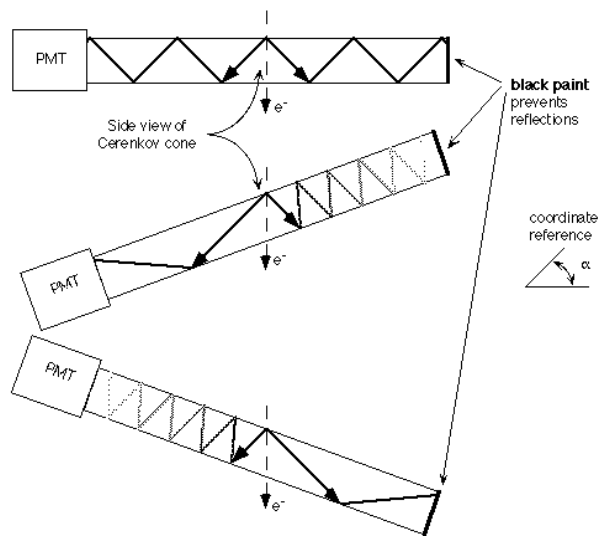

Fig. 5
Brian Porter
*American Journal of Physics*

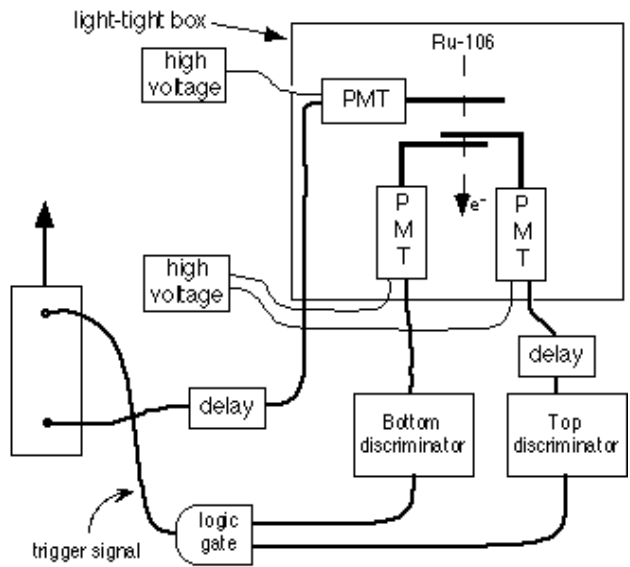

Fig. 6
Brian Porter
*American Journal of Physics*

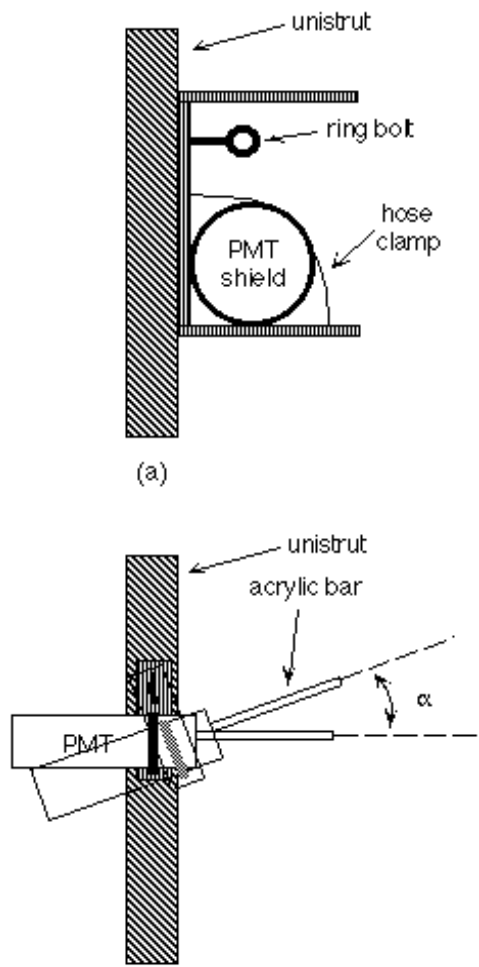

Fig. 7
Brian Porter
*American Journal of Physics*
22

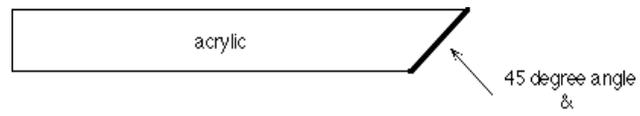

Fig. 8
Brian Porter
*American Journal of Physics*

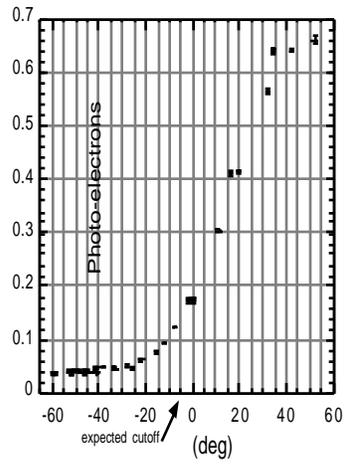

Fig. 9
Brian Porter
*American Journal of Physics*

24